\documentclass[twocolumn,showpacs,aps,prl,superscriptaddress]{revtex4}
\usepackage{graphicx}
\usepackage{amsmath}
\usepackage{epsfig}
\newcommand{\BABARPubYear}    {09}
\newcommand{\BABARPubNumber}  {012}
\newcommand{\SLACPubNumber} {13711}

\def\too	{\ensuremath{\!\!\to}}
\def\faz	{\ensuremath{f_A\!=0}\xspace}
\def\fafv	{\ensuremath{f_A\!=\!f_V}\xspace}
\def\enugam     {\ensuremath{\Bp\too e^+\nue\g}\xspace }
\def\mnugam	{\ensuremath{\Bp\too\mu^+\num\g}\xspace }
\def\lnugam	{\ensuremath{\Bp\too\ellp\nul\g}\xspace }
\def\pilnu	{\ensuremath{\Bp\too\piz\ellp\nul}\xspace} 
\def\etalnu	{\ensuremath{\Bp\too\eta\ellp\nul}\xspace} 
\def\etaPlnu	{\ensuremath{\Bp\too\eta^{\prime}\ellp\nul}\xspace} 
\def\Xlnu	{\ensuremath{\Bp\too X_u^0 \ellp\nul}\xspace}  
\def\Btag	{\ensuremath{B_{\rm tag}}\xspace}  
\def\nuMass	{\ensuremath{m_{\nu}^2}\xspace}  
\def\Npeak	{\ensuremath{N_{\ell}^{\rm peak}}\xspace}  
\def\Ncomb	{\ensuremath{N_{\ell}^{\rm comb}}\xspace}  
\def\Nbkg	{\ensuremath{N_{\ell}^{\rm bkg}}\xspace}  
\def\eff	{\ensuremath{\eps_{\ell}^{\rm sig}}\xspace} 

\input babarsym

\def\figurebox#1#2#3{%
    \def\arg{#3}%
    \ifx\arg\empty
    {\hfill\vbox{\hsize#2\hrule\hbox to #2{\vrule\hfill\vbox to #1{\hsize#2\vfill}\vrule}\hrule}\hfill}%
    \else
    {\hfill\epsfbox{#3}\hfill}%
    \fi}

\begin{document}

\begin{flushleft}
\babar-PUB-\BABARPubYear/\BABARPubNumber \\
SLAC-PUB-\SLACPubNumber  \\[10mm]
\end{flushleft}

\title{\large \bf \boldmath Model-independent search for the decay \lnugam}
%% author list as of 03-Apr-2009 (488 authors)
%
\author{B.~Aubert}
\author{Y.~Karyotakis}
\author{J.~P.~Lees}
\author{V.~Poireau}
\author{E.~Prencipe}
\author{X.~Prudent}
\author{V.~Tisserand}
\affiliation{Laboratoire d'Annecy-le-Vieux de Physique des Particules (LAPP), Universit\'e de Savoie, CNRS/IN2P3,  F-74941 Annecy-Le-Vieux, France}
\author{J.~Garra~Tico}
\author{E.~Grauges}
\affiliation{Universitat de Barcelona, Facultat de Fisica, Departament ECM, E-08028 Barcelona, Spain }
\author{M.~Martinelli$^{ab}$}
\author{A.~Palano$^{ab}$ }
\author{M.~Pappagallo$^{ab}$ }
\affiliation{INFN Sezione di Bari$^{a}$; Dipartimento di Fisica, Universit\`a di Bari$^{b}$, I-70126 Bari, Italy }
% ...
\collaboration{The \babar\ Collaboration}
\noaffiliation

\begin{abstract}
We present a search for the radiative leptonic decay \lnugam, where $\ell=e,\mu$, using a data sample of $465\times10^{6}$ \BB pairs collected by the \babar\ experiment.  In this analysis, we fully reconstruct the hadronic decay of one of the $B$ mesons in $\FourS\to\BpBm$ decays, then search for evidence of \lnugam in the rest of the event.  We observe no significant evidence of signal decays and report model-independent branching fraction upper limits of $\BR(\enugam)<17\times10^{-6}$, $\BR(\mnugam)< 24\times10^{-6}$, and $\BR(\lnugam)<15.6\times10^{-6}$ ($\ell=e~{\rm or}~\mu$), all at the 90\% confidence level. 
\end{abstract}

\maketitle

The leptonic decay \lnugam \cite{ref:ccNote}, where $\ell= e~{\rm or}~\mu$, proceeds via quark annihilation into a virtual $W^+$ boson with the radiation of a photon.  The presence of the photon removes the helicity suppression of the purely leptonic decays, $B^+ \too \ell^+ \nu_{\ell}$, although it introduces an additional suppression by a factor of $\alpha_{\rm em}$.  The branching fraction of \lnugam is predicted to be of order $10^{-6}$ \cite{ref:BF}, making it potentially accessible at $B$ factories.  The most stringent published limits are from the CLEO Collaboration with \BR(\enugam) $< 2.0\times 10^{-4}$ and \BR(\mnugam) $< 5.2 \times 10^{-5}$ at the 90\% confidence level (C.L.) \cite{ref:cleo}.

The differential branching fraction versus photon energy $E_{\g}$ involves two form factors, $f_V$ and $f_A$, which contain the long-distance contribution of the vector and axial currents, respectively, in the $\B\to\g$ transition
\begin{equation}
	\label{diffBF}
	\frac{d\BR}{dE_{\g}} = \frac{\alpha_{\rm em} G_F^2\lvert V_{ub}\rvert^2}{48\pi^2}m_B^4\tau_B\left[f_A^2(E_{\g})+f_V^2(E_{\g})\right](1-y)y^3 ,
\end{equation}
where $G_F$ is the Fermi constant, $V_{ub}$ is the Cabibbo-Kobayashi-Maskawa quark-mixing matrix element describing the coupling of $b$ and $u$ quarks, $m_B$ and $\tau_B$ are the $B$-meson mass and lifetime, respectively, and ${y\equiv2E_{\g}/m_B}$.  While \fafv in most models \cite{ref:Lunghi}, some suggest \faz \cite{ref:burdman}.  The branching fraction is given by Ref.~\cite{ref:KPY} as
\begin{equation}
	\label{lnugamBF}
	\BR(\lnugam) = \frac{\alpha_{\rm em} G_F^2\lvert V_{ub}\rvert^2}{288\pi^2}f^2_Bm_B^5\tau_B\left(\frac{Q_u}{\lambda_B}-\frac{Q_b}{m_b}\right)^2 ,
\end{equation}
where $f_B$ is the $B$-meson decay constant, $Q_{u,b}$ are the $u$- and $b$-quark charges, and $m_b$ is the $b$-quark mass.  The first inverse moment of the $B$-meson distribution amplitude ${\lambda}_B$ is expected to be of order $\Lambda_{\rm QCD}$ but its theoretical estimation suffers from large uncertainties \cite{ref:ball}.  It also appears in the branching fractions of two-body hadronic $B$-meson decays, such as ${\B\to \pi \pi}$, and plays an important role in QCD factorization \cite{ref:Lunghi}.  Since there are no hadrons in the final state, an experimental measurement of \lnugam can provide a clean determination of ${\lambda}_B$.

We present the first search for \lnugam that exploits the hadronic ``recoil'' technique, in which one $B$ meson is exclusively reconstructed in a hadronic final state before searching for the signal decay within the rest of the event.  This technique improves the handling of event kinematics,  providing adequate background suppression without requiring model-dependent constraints on the signal kinematics.  Thus, this analysis is valid for all $B\to\g$ form-factor models and over the full kinematic range.  This analysis uses a data sample of $465 \pm 5$ million \BB pairs, corresponding to an integrated luminosity of $423\,\invfb$ collected at the \FourS resonance.  The data were recorded with the \babar\ detector at the asymmetric-energy \pep2 $e^+e^-$ storage ring at SLAC.  The \babar\ detector is described in detail elsewhere \cite{ref:babar}. 

Signal and background decays are studied using Monte Carlo (MC) samples based on GEANT4 \cite{ref:geant4}.  The simulation includes a detailed model of the \babar\ detector geometry and response.  Beam-related background and detector noise are extracted from data and overlaid on the MC simulations.  $\FourS\to\BB$ signal MC samples are generated with one $B$ meson decaying via \lnugam using the tree-level model of Ref.~\cite{ref:KPY}, which is valid for $y>0.13$, while the other $B$ meson decays generically.  We simulate signal MC samples for two form-factor models, with \fafv and \faz respectively, to evaluate the impact of the decay model on the signal selection efficiency.  Large MC samples of generic \BB and continuum ($e^+e^-\too\tautau$ or $e^+e^-\too\qqbar$, where $q = u,d,s,c$) events are used to optimize the signal selection criteria.  However, the final background estimates are obtained directly from a combination of data and exclusive \Xlnu MC samples, where $X_u^0$ is a neutral meson containing a $u$ quark.  The primary background for \lnugam in this analysis is due to \Xlnu decays, with \pilnu (\etalnu) comprising approximately 73\% (18\%) of this semileptonic background.  The branching fraction and uncertainty for each \Xlnu mode are taken from experimental measurements ($X_u^0$ = \piz \cite{ref:pdg}, $\rho$ \cite{ref:pdg}, $\eta$ \cite{ref:etaBF}, and $\omega$ \cite{ref:omegaBF}).  We assume \BR(\etaPlnu) = $\BR(\etalnu)\times(1\pm1)$.  We use a light-cone sum rule model for the $\eta$ and $\eta^{\prime}$ form factors \cite{ref:ball04} and use the form factor measured in a \babar\ analysis \cite{ref:cote} with the shape parameterization given in Ref.~\cite{ref:coteShape} for the \piz mode.

Event selection begins with the full reconstruction of a charged $B$ meson (\Btag) in one of the large number of hadronic final states, $\Bm\too \Db^{(*)}X_{\rm had}$.  We reconstruct the $D^{*-}\too \Dzb\pi^-$; $\Dstarzb\too \Dzb\piz$, $\Dzb\g$; $D^-\too\KS\pi^-$, $\KS\pi^-\piz$, $\KS\pi^-\pi^-\pi^+$, $ K^+\pi^-\pi^-$, $K^+\pi^-\pi^-\piz$; $\Dzb\too K^+\pi^-$, $K^+\pi^-\piz$, $K^+\pi^-\pi^-\pi^+$, $\KS\pi^-\pi^+$; and $\KS\too\pi^-\pi^+$ decay modes.  $X_{\rm had}$ is a collection of at most five mesons, composed of both charged and neutral kaons and pions.  Well-reconstructed \Btag candidates are selected using two kinematic variables: $\Delta E= E_{\Btag}-\sqrt{s}/2$ and $\mes = \sqrt{s/4 - {\vec p}^{\,2}_{\Btag}}$, where $E_{\Btag}$ and ${\vec p}_{\Btag}$ are the energy and momentum of the \Btag candidate, respectively, and $\sqrt{s}$ is the total energy of the \epem system, all in the center-of-mass (CM) frame.  We require $\Delta E$, which peaks at zero for correctly reconstructed $B$ mesons, to lie between $-0.12$ and $0.12\gev$ or within two standard-deviations from its mean for the given $X_{\rm had}$ mode, whichever is the tighter constraint.  We fit the \mes distribution for each $X_{\rm had}$ mode and require that the purity, or fraction of well-reconstructed $B$ mesons, is greater than 12\% in the region $\mes>5.27\gevcc$.  If more than one \Btag candidate is reconstructed, the one in the highest purity mode is chosen.  If there are multiple candidates in this mode, the one that minimizes $\lvert\Delta E\rvert$ is selected.

\begin{figure}
 \includegraphics[width=3.4in]{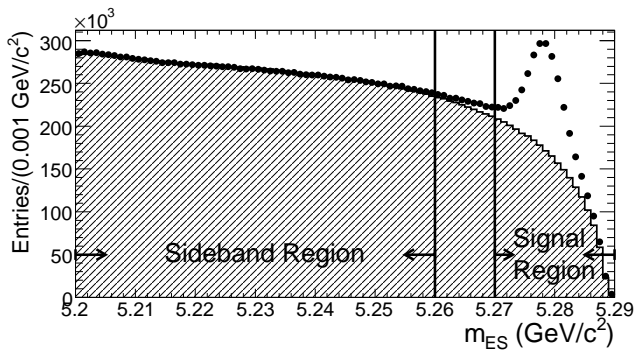}
 \caption{\mes distribution, after \Btag reconstruction and continuum suppression, of data (points) and the expected combinatoric background as predicted by the MC (shaded). \label{fig:mes}} 
\end{figure}

We define the signal region as $5.27<\mes<5.29\linebreak[1]\gevcc$, since correctly reconstructed $B$ mesons peak in this region near the nominal $B$-meson mass.  The \Btag candidates that are incorrectly reconstructed from either continuum events or both $B$ mesons (``combinatoric" events), produce a distribution that is fairly flat below the \mes signal region and decreases within it, as shown in Fig.~\ref{fig:mes}.  The shape of the combinatoric distribution is extrapolated into the \mes signal region using MC, while the background contribution from combinatoric events is estimated directly from the data.  To improve the MC estimate of the \Btag reconstruction efficiency,  we normalize the generic MC to the number of data events that peak within the \mes signal region.  Thus, all MC samples are scaled by 90.7\%, resulting in good agreement between data and background MC throughout the analysis selection.  A charged \Btag is reconstructed in about 0.3\% of the signal MC events.

Because the two $B$ mesons produced in the \FourS decay have low momenta in the CM frame ($0.3\gevc$), their decay products are more isotropic than continuum background.  For example, $\lvert\cos\theta_{\rm T}\rvert$, where $\theta_{\rm T}$ is the angle in the CM frame between the \Btag thrust axis and the thrust axis of all other particles in the event, has a flat distribution for \BB events and peaks near one for non-\BB events.  The continuum background is suppressed by requiring $\calL_{B}\equiv \prod_i {\cal P}_{B}(x_i)/\left(\prod_i {\cal P}_{B}(x_i)+\prod_i {\cal P}_{q}(x_i)\right)>30\%$, where ${\cal P}_{B}(x_i)$ (${\cal P}_{q}(x_i)$) are probability density functions determined from MC that describe \BB (continuum) events for the five event-shape variables $x_i$.  The variables used are: the ratio of the second to zeroth Fox-Wolfram moment \cite{ref:foxWolf} computed using all charged and neutral particles in the event, the cosine of the angle between ${\vec p}_{\Btag}$ and the beam axis, the magnitude of the \Btag thrust, the component of the \Btag thrust along the beam axis, and $\lvert\cos\theta_{\rm T}\rvert$.  This requirement improves the agreement between data and MC by suppressing unmodeled continuum backgrounds, such as $e^+e^-\too e^+e^-\ell^+\ell^-$ via two photons.

In the sample of selected \Btag candidates, we identify events in which the remaining tracks, calorimeter clusters, and missing momentum vector (${\vec p}_{\rm miss}$) are consistent with \lnugam candidates.  We select events with exactly one track, which reduces the signal efficiency by 25\% but removes over $99\%$ of the simulated background events with a reconstructed \Btag.  This signal track is required to have a charge opposite to that of the \Btag, to satisfy particle identification (PID) criteria for either a muon or electron, and to be inconsistent with a kaon hypothesis.  In the electron mode, the four-momenta of signal tracks are redefined to include those of any bremsstrahlung photon candidates.  Such a candidate is defined as any cluster whose momentum vector, when compared to that of the signal track (${\vec p}_{\ell}$), is separated by $\lvert\Delta\theta\rvert<3^{\circ}$ and $-3^{\circ}<Q_e\times \Delta\phi<13^{\circ}$, where $Q_e=\pm1$ is the $e^\pm$ charge and $\theta$ ($\phi$) is the polar (azimuthal) angle relative to the beam axis, in the lab frame.  Finally, the signal photon candidate is chosen as the cluster with the highest CM energy, excepting bremsstrahlung photon candidates.

\begin{figure}
 \includegraphics[width=3.4in]{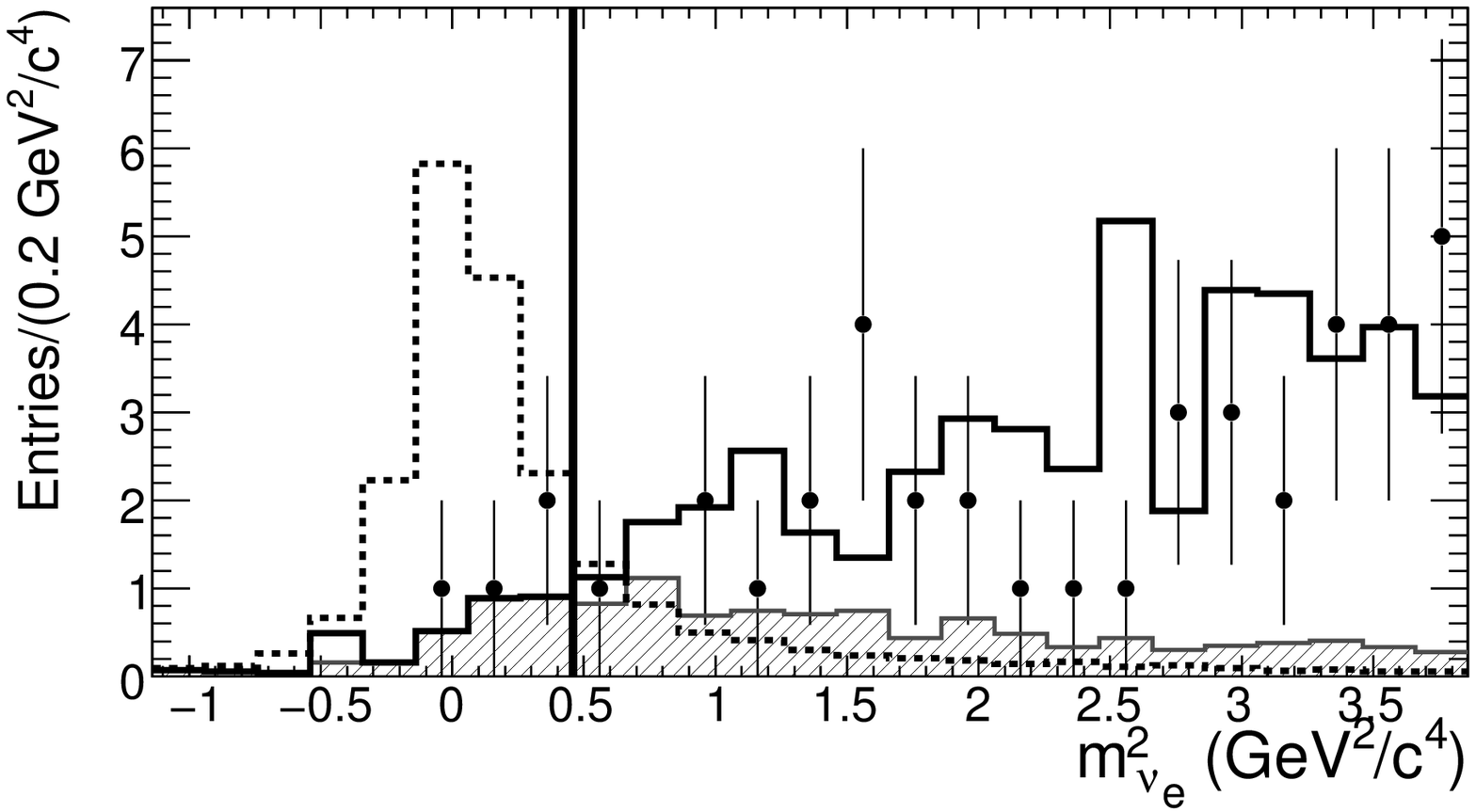}
 \\
 \includegraphics[width=3.4in]{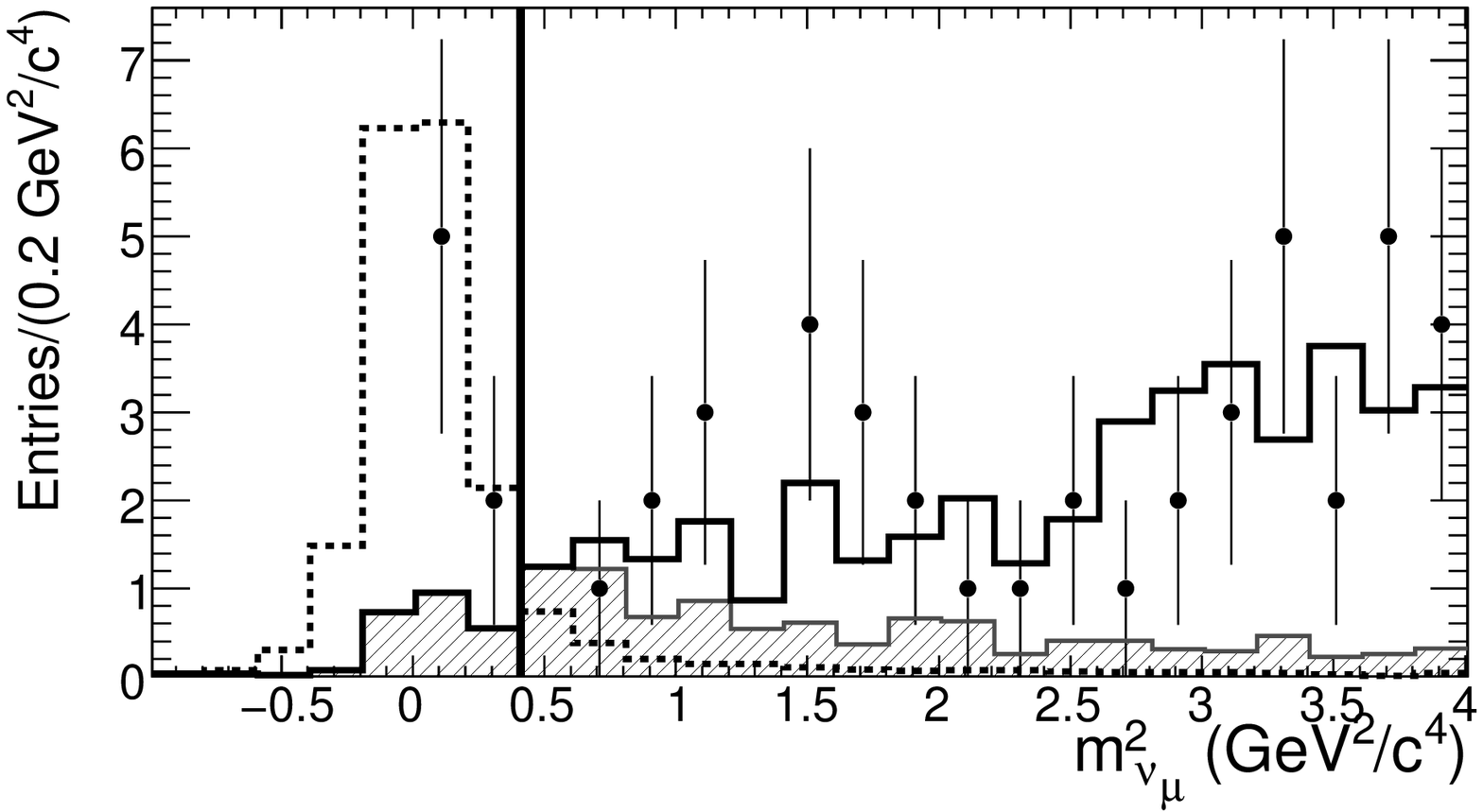}
 \caption{\nuMass distribution after all selection criteria are applied, in electron (top) and muon (bottom) modes for the \mes-peaking (shaded) plus nonpeaking (solid) contributions in the full background MC sample, signal MC normalized to $\BR=40\times10^{-6}$ (dashed), and data (points).  Events to the left of the vertical lines are selected.  \Ncomb of Table~\ref{tab:finalNums} is determined from sideband data, not from the MC shown here.  \label{fig:nuMass}} 
\end{figure}
We significantly reduce the background by requiring that the kinematics of the signal track and photon candidate are consistent with the existence of a third massless particle originating from the signal $B$ meson.  To do this, we use the four-momentum of the expected signal $B$ meson ($p_{B}$), which is assigned an energy of $\sqrt{s}/2$, a momentum vector pointing along $-\vec{p}_{\Btag}$, and the nominal $B$-meson mass.  The neutrino mass squared is then defined as $\nuMass \equiv ( p_{B}-p_{\ell}-p_{\g})^2$, where $p_{\ell}$ ($p_{\g}$) is the four-momentum of the signal track (photon candidate).  As shown in Fig.~\ref{fig:nuMass}, the background increases with \nuMass, while \lnugam events peak at $\nuMass=0$ with an enhanced tail in the electron mode due to unrecovered bremsstruhlung photons.  We require $-1<m_{\nu}^2<0.46~(0.41)\mathrm{\,Ge\kern -0.1em V^2\!/}c^4$ for the electron (muon) modes.  In addition, the lepton and neutrino should be emitted back-to-back in the rest frame that recoils from the photon emission, defined as $p_{B}-p_{\g}$.  We require $\cos\theta_{\ell\nu}<-0.93$ in this frame, where $\theta_{\ell\nu}$ is the angle between ${\vec p}_{\ell}$ and ${\vec p}_{\rm miss}$.  After all other selection criteria are applied, the MC indicates that \nuMass and $\cos\theta_{\ell\nu}$ together remove 99\% of background events with a 30 and 20\% reduction in the signal efficiency for the electron and muon modes, respectively.

The dominant backgrounds are due to $\Bp\too \piz\ellp\nul$ $(\eta\ellp\nul)$ events in which $\piz (\eta) \!\to \g\g$ fakes the \lnugam signal photon.  To suppress this background, we reject events containing a $\piz (\eta)$ candidate, reconstructed using the signal photon candidate and a second cluster having CM energy $E_{\g_2}$. For \piz candidates, we require a $\g\g$ invariant mass between 120--145\mevcc with $E_{\g_2}>30\mev$ or between 100--160\mevcc with $E_{\g_2}>80\mev$.  For $\eta$ candidates, we require a $\g\g$ invariant mass between 515--570\mevcc with $E_{\g_2}>100\mev$.  Likewise, $\Bp\too\omega\ellp\nul\!\to\![\piz\g]\ellp\nul$ events are suppressed by rejecting any event in which the signal photon candidate and a \piz candidate produce an invariant mass between 730--830 \mevcc.  This \piz candidate is defined as any two clusters with CM energy $>70\mev$ which produce a $\g\g$ invariant mass between 115--145\mevcc.  After applying all other selection criteria, these vetoes reduce the \pilnu and \Xlnu background events, with $X_u^0\ne\piz$, by 65\% and 50\% respectively. Finally, we require the lateral moment \cite{ref:latMom} of the calorimeter energy deposit for the signal photon candidate, which peaks at 25\% for single photons, to be between 0 and 55\%.  This suppresses \pilnu events in which the two photons from the \piz decay are reconstructed as a single merged photon.

Once the \Btag, signal photon, and lepton are identified, \lnugam events are expected to contain little or no additional energy within the calorimeter.  However, additional energy deposits can result from hadronic shower fragments, beam-related photons, and photons from unreconstructed $\Dstarb\too\Db\g/\piz$ transitions in the \Btag candidate.  The total energy of all additional clusters is required to be less than $0.8\gev$, counting only clusters with lab-frame energy greater than $50\mev$.  We also require that ${\vec p}_{\rm miss}$ points within the fiducial acceptance of the detector.

To avoid experimenter bias, we optimize all the selection criteria and determine the number of expected background events in the signal region (\Nbkg), for $\ell = e$ or $\mu$, before looking at any data events selected by the criteria.  We optimize by maximizing the figure of merit $\eff/(\tfrac{1}{2}n_{\sigma}+\sqrt{\Nbkg})$ \cite{ref:punzi}, where $n_{\sigma} = 1.3$ and \eff is the total signal efficiency including that of the \Btag reconstruction.  The signal branching fraction is calculated using $\BR_{\ell} = (N_{\ell}^{\rm obs}-\Nbkg)/\eff N_{B^{\pm}}$, where $N_{B^{\pm}}=465\times10^{6}$ is the number of $B^{\pm}$ mesons in the data sample and $N_{\ell}^{\rm obs}$ is the number of data events within the signal region.

\begin{figure}
 \includegraphics[width=3.4in]{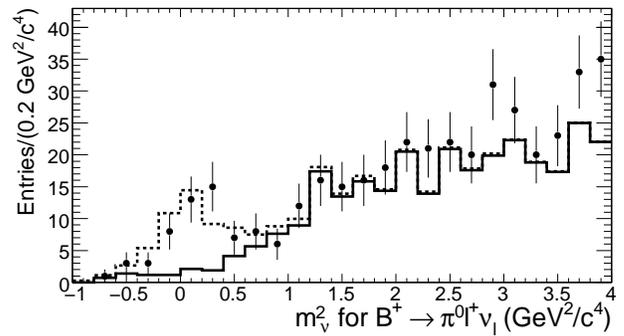}
 \caption{\nuMass distribution for \pilnu ($\ell=e~{\rm and}~\mu$), using the procedure described in the text where \g is substituted with a \piz candidate, of data (points) and of \pilnu MC normalized to $\BR=7.7\times10^{-5}$ (dashed) and added to the expected background (solid).
  \label{fig:control}} 
\end{figure}

To verify the modeling of \eff, we remove the \Xlnu vetoes, select events containing a \piz candidate, and substitute the \piz in place of the signal photon candidate.  The resulting \nuMass distribution from \pilnu is expected to resemble that of the signal. We observe a peak in the data that agrees with MC expectations within the 15\% statistical uncertainty of the data, as shown in Fig.~\ref{fig:control}.  For cross-check purposes only, we determine the \pilnu efficiency using an exclusive \pilnu MC sample and the background contribution using generic MC.  The peak in data corresponds to $\BR(\pilnu) = (7.8 _{-1.1}^{+1.7})\times 10^{-5}$, where the uncertainty is statistical.  This branching fraction is consistent with the current world-average value of $(7.7\pm1.2)\times 10^{-5}$ \cite{ref:pdg}, which is also the value used in the MC samples.

The total number of background events \Nbkg has two components: \Npeak the number of expected background events having a correctly reconstructed \Btag and hence peaking within the \mes signal region, and \Ncomb the number of expected combinatoric background events, including both \BB and continuum events.  The \nuMass and $\cos\theta_{\ell\nu}$ restrictions ensure kinematic and topological consistency with a three-body decay involving a massless and undetected particle: the neutrino.  By further requiring that exactly one track recoils from a fully-reconstructed \Btag, lepton number and PID ensures the track is a lepton.  Thus, only \lnugam decays can peak within the signal region, unless the signal photon candidate actually arises from one or more particles that mimic the kinematics of \lnugam, which only occurs in specific pathological \Xlnu decays.  Therefore, we determine \Npeak using exclusive \Xlnu MC simulations and validate the lack of additional peaking backgrounds with generic MC.  Other decay modes passing the selection criteria do so with poorly reconstructed \Btag candidates and thus produce a combinatoric distribution in \mes.  We determine \Ncomb from an extrapolation of the observed number of data events within the \mes sideband region, defined as $5.20<\mes<5.26\gevcc$.  We observe 1 (4) data events within the \mes sideband for the electron (muon) mode.

The uncertainty on \Ncomb is dominated by the sideband data statistics.  It also includes the systematic uncertainty from the combinatoric background shape, estimated by varying the selection criteria and the method used to extrapolate this shape (14.6\%).  The error on \Npeak is dominated by uncertainties in the branching fractions and form factors associated with the various exclusive \Xlnu decays (13.6\%).  Additional systematic uncertainties result from MC modeling of the data efficiency, which we apply to both \Npeak and \eff: electron PID (0.9\%) or muon PID (1.3\%), $\calL_{B}$ (1.4\%), \nuMass (0.5\% for \eff, 1.4\% for \Npeak), and the reconstructions of the track (0.4\%), photon (1.8\%), and \Btag (3.1\%).  The last of these, which also accounts for uncertainty in $N_{B^{\pm}}$, is estimated by varying the shape of the \mes combinatoric distribution and the size of the \mes signal and sideband regions.

Branching fraction limits and uncertainties are computed using the frequentist formalism of Feldman and Cousins \cite{ref:FC}, with the uncertainties on \Nbkg and \eff modeled using Gaussian distributions.  Since \BR(\lnugam) is expected to be independent of the lepton type, we also combine the two modes by maximizing a likelihood function defined as the product of both Poisson probabilities in $N_{\ell}^{\rm bkg}$, where $\BR_{\ell}$ is the mean.

We observe 4 (7) data events within the signal region for the electron (muon) mode, compared to an expected background of 2.7 $\pm$ 0.6 (3.4 $\pm$ 0.9) events.  This corresponds to a signal significance of $1.2\sigma~(1.8\sigma)$, a combined significance of $2.1\sigma$, and the results given in Table~\ref{tab:finalNums}.  The effective detector and PID thresholds are about 20\mev for photon energy and $400~(800)\mevc$ for electron (muon) momentum, and we apply no minimum energy requirements.  Thus, this analysis is essentially independent of the kinematic model; we assume the \fafv signal model, but the \faz model yields consistent \eff values.  Since certain theoretical calculations are most reliable at high $E_{\g}$ \cite{ref:ball}, we also report a partial branching fraction limit $\Delta\BR$ by selecting events with a photon candidate energy greater than 1\gev, which reduces \eff by 30\%.  We observe 2 (4) data events with $\Nbkg=1.4 \pm 0.3~(2.5 \pm 1.0)$, resulting in $\Delta\BR(\lnugam)<14\times10^{-6}$ at 90\% C.L.  

\begin{table*}
	 \caption{Expected background yields $\Nbkg\!=\Ncomb+\Npeak$, signal efficiencies \eff, number of observed data events $N_{\ell}^{\rm obs}$, resulting branching fraction limits at 90\% C.L., and the combined central value $\BR_{\rm combined}$.  Model-specific limits are also presented.  Uncertainties are given as statistical $\pm$ systematic. \label{tab:finalNums}}
	\begin{ruledtabular}
 	\begin{tabular}{lccc}
           & \enugam                                     & \mnugam                             & \lnugam \\ \hline 
\Ncomb     &  0.3 $\pm$  0.3 $\pm$ 0.1                   &  1.2 $\pm$  0.6 $\pm$  0.6          & \\
\Npeak     &  2.4 $\pm$  0.3 $\pm$  0.4                  &  2.1 $\pm$  0.3 $\pm$  0.3          & \\ 
\Nbkg      &  2.7 $\pm$  0.4 $\pm$  0.4                  &  3.4 $\pm$  0.7 $\pm$  0.7          & \\
\eff       & (7.8 $\pm$  0.1 $\pm$  0.3)$\times10^{-4}$  & (8.1 $\pm$  0.1 $\pm$  0.3)$\times10^{-4}$ & \\
$N_{\ell}^{\rm obs}$     & 4                             & 7                                   & \\ 
$\BR_{\rm combined}$         &       &        & $\bigl(6.5 _{-4.7}^{+7.6}$$_{-0.8}^{+2.8}\bigr)\times10^{-6}$ \\ 
Model-independent limits & $< 17$$\times10^{-6}$         & $< 26$$\times10^{-6}$               & $<15.6\times10^{-6}$ \\ 
\fafv limits             & $<8.4\times10^{-6}$           & $<6.7\times10^{-6}$                 & $<3.0\times10^{-6}$ \\
\faz limits              & $<29\times10^{-6}$            & $<22\times10^{-6}$                  & $<18\times10^{-6}$ \\
 	\end{tabular}
	\end{ruledtabular}
\end{table*}

In Table~\ref{tab:finalNums}, we also report model-specific limits by introducing a kinematic requirement on the relationship between $\cos\theta_{\g\ell}$ and $\cos\theta_{\g\nu}$, where $\theta_{\g\ell}$ ($\theta_{\g\nu}$) is the angle between the photon candidate momentum and ${\vec p}_{\ell}$ (${\vec p}_{\rm miss}$) in the signal $B$ rest frame.  The photon is emitted preferentially back-to-back with the lepton in the \fafv model, and with either the lepton or neutrino when \faz.  Thus, we require $(\cos\theta_{\g\ell}-1)^2 +(\cos\theta_{\g\nu}+1)^2/3 >0.4$ or $(\cos\theta_{\g\nu}-1)^2 +(\cos\theta_{\g\ell}+1)^2/3 >0.4$ for the \faz~model, and only the former relationship for \fafv.  This reduces \eff in both modes and models by 40\%.  We observe 0 (0) data events in the electron (muon) mode with \Nbkg $= 0.6 \pm 0.1~(1.0 \pm 0.4)$ for the \fafv model, and 3 (2) data events with $\Nbkg = 1.2 \pm 0.4~(1.5 \pm 0.6)$ for \faz.  

In conclusion, we have searched for \lnugam using a hadronic recoil technique and observe no significant signal within a data sample of $465\times10^6$ \BB pairs.  We present model-specific branching fraction limits in Table~\ref{tab:finalNums}.  We also report a model-independent limit of $\BR(\lnugam)<15.6\times10^{-6}$ at the 90\% C.L., which is consistent with the standard model prediction and is the most stringent published upper limit to date.  Using Eq.\eqref{lnugamBF} with $f_B = 0.216\pm0.022\gev$ \cite{ref:fBtheory}, $m_B=5.279\gevcc$, $\tau_B=1.638\,\ps$, $m_b = 4.20\gevcc$, and $\lvert V_{ub}\rvert = (3.93\pm0.36)\times10^{-3}$ \cite{ref:pdg}, the combined branching fraction likelihood function corresponds to a limit of $\lambda_B\!> 0.3\gev$ at the 90\% C.L.

We are grateful for the excellent luminosity and machine conditions
provided by our \pep2\ colleagues, 
and for the substantial dedicated effort from
the computing organizations that support \babar.
The collaborating institutions wish to thank 
SLAC for its support and kind hospitality. 
This work is supported by the
DOE
and NSF (USA),
NSERC (Canada),
CEA and
CNRS-IN2P3
(France),
BMBF and DFG
(Germany),
INFN (Italy),
FOM (The Netherlands),
NFR (Norway),
MES (Russia),
MEC (Spain), and
STFC (United Kingdom). 
Individuals have received support from the
Marie Curie EIF (European Union) and
the A.~P.~Sloan Foundation.


\begin{thebibliography}{99}
\bibitem{ref:ccNote} {Charge conjugation is implied throughout this paper.}
\end{thebibliography}
\end{document}